\documentclass[12pt]{article}%
\usepackage{amsmath}
\usepackage{amsfonts}
\usepackage{amssymb}
\usepackage{graphicx}%
\begin{document}

\title{Quantum spherical model with competing interactions}
\author{P. F. Bienzobaz and S. R. Salinas\\Instituto de F\'{\i}sica, Universidade de S\~{a}o Paulo,\\Caixa Postal 66318\\05314-970, S\~{a}o Paulo, SP, Brazil}
\date{\today }
\maketitle

\begin{abstract}
We analyse\ the phase diagram of a quantum mean spherical model in terms of
the temperature $T$, a quantum parameter $g$, and the ratio $p=-J_{2}/J_{1}$,
where $J_{1}>0$ refers to ferromagnetic interactions between first-neighbour
sites along the $d$ directions of a hypercubic lattice, and $J_{2}<0$ is
associated with competing antiferromagnetic interactions between second
neighbours along $m\leq d$ directions. We regain a number of known results for
the classical version of this model, including the topology of the critical
line in the $g=0$ space, with a Lifshitz point at $p=1/4$, for $d>2$, and
closed-form expressions for the decay of the pair correlations in one
dimension. In the $T=0$ phase diagram, there is a critical border,
$g_{c}=g_{c}\left(  p\right)  $ for $d\geq2$, with a singularity at the
Lifshitz point if $d<(m+4)/2$. We also establish upper and lower critical
dimensions, and analyse the quantum critical behavior in the neighborhood of
$p=1/4$.

\end{abstract}

\section{Introduction}

The spherical model of magnetism has been used as an excellent laboratory to
test ideas and concepts of phase transitions and critical phenomena
\cite{Berlin1952}\cite{Lewis1952}\cite{Joyce1972}. There are several versions
of the original model, including proposals of a quantum spherical model to
correct some of the unphysical results at low temperatures \cite{Obermair1972}%
\cite{Nieuwenhuizen1995}\cite{Vojta1996}\cite{Coutinho2005}. The effects of
frustration \cite{Chayes1996}\cite{Nussinov2004}, random fields
\cite{Vojta.Schreiber1996}, and of disordered exchange interactions
\cite{Bolle2003}\cite{Menezes.Theumann2007}, have also been analysed in the
context of quantum spherical models. With a view to describe the crossover
between classical and quantum critical behaviour, Vojta \cite{Vojta1996} used
a standard scheme of canonical quantization to analyse a quantum version of
the ferromagnetic mean spherical model. We were then motivated to revisit this
problem, with the addition of competing ferro and antiferromagnetic
interactions, and the perspective to analyse a quantum Lifshitz point.

The mean spherical model, which has been originally proposed by Lewis and
Wannier \cite{Lewis1952}, is given by the partition function
\begin{equation}
Z_{cl}=\left[  {\prod\limits_{\overrightarrow{l}}\int\limits_{-\infty
}^{+\infty}}dS_{\overrightarrow{l}}\right]  \exp\left[  -\beta\,\mathcal{H}%
\left\{  S_{\overrightarrow{l}}\right\}  -\beta\mu\,{\sum
\limits_{\overrightarrow{l}}}S_{\overrightarrow{l}}^{2}\right]  ,
\end{equation}
where $\beta=1/\left(  k_{B}T\right)  $, $T$ is the temperature and $k_{B}$ is
the Boltzmann constant, $\mu$ is a suitable chemical potential,
$\overrightarrow{l}$ is a lattice vector, and $\left\{  S_{\overrightarrow{l}%
}\right\}  $ is a set of continuous spin variables running over the $N^{d}$
sites of a $d$-dimensional hypercubic lattice. The model Hamiltonian is
written as%
\begin{equation}
\mathcal{H}=-\sum_{\left(  \overrightarrow{k},\overrightarrow{l}\right)
}J_{\overrightarrow{k},\overrightarrow{l}}~S_{\overrightarrow{k}%
}S_{\overrightarrow{l}}-H\sum_{\overrightarrow{l}}S_{\overrightarrow{l}},
\end{equation}
where $\left(  \overrightarrow{k},\overrightarrow{l}\right)  $ labels a pair
of lattice sites, the exchange parameter $J_{\overrightarrow{k}%
,\overrightarrow{l}}=J\left(  \left\vert \overrightarrow{k}-\overrightarrow
{l}\right\vert \right)  $ depends on the distance between sites
$\overrightarrow{k}$ and $\overrightarrow{l}$, and $H$ is an external field.
In this formulation, the chemical potential $\mu$\ comes from the mean
spherical condition,%
\begin{equation}
\left\langle
{\displaystyle\sum\limits_{\overrightarrow{l}}}
S_{\overrightarrow{l}}^{2}\right\rangle =-\frac{1}{\beta}\frac{\partial
}{\partial\mu}\ln Z_{cl}=N^{d},\label{sphconstr}%
\end{equation}
and it is well known that exact solutions for the thermodynamic functions can
be obtained from the standard diagonalization of a quadratic form
\cite{Joyce1972}.

In a quantum version of this mean spherical model \cite{Obermair1972}%
\cite{Vojta1996}, the spin variable $S_{\overrightarrow{l}}$ becomes a
position operator at lattice site $\overrightarrow{l}$, canonically conjugate
to a momentum operator $P_{\overrightarrow{l}}$, with the commutation
relations%
\begin{equation}
\lbrack S_{\overrightarrow{l}},S_{\overrightarrow{k}}]=0,\quad\lbrack
P_{\overrightarrow{l}},P_{\overrightarrow{k}}]=0,\quad\lbrack
S_{\overrightarrow{l}},P_{\overrightarrow{k}}]=i\delta_{\overrightarrow
{l},\overrightarrow{k}},
\end{equation}
where $\delta_{\overrightarrow{l},\overrightarrow{k}}$ is a Kronecker delta
and we assume that $\hbar=1$. We then add a term of kinetic energy, depending
on a quantum parameter $g$, and write the quantum quadratic form%
\begin{equation}
\overline{\mathcal{H}}=\frac{1}{2}g\sum_{\overrightarrow{l}}P_{\overrightarrow
{l}}^{2}-\sum_{\left(  \overrightarrow{k},\overrightarrow{l}\right)
}J_{\overrightarrow{k},\overrightarrow{l}}~S_{\overrightarrow{k}%
}S_{_{\overrightarrow{l}}}-H\sum_{\overrightarrow{l}}S_{\overrightarrow{l}%
}+\mu%
{\displaystyle\sum\limits_{\overrightarrow{l}}}
S_{\overrightarrow{l}}^{2},\label{hquantum}%
\end{equation}
which can be diagonalized by a canonical method \cite{Vojta1996}, leading to a
solution of the problem for a general ferromagnetic pair interaction. At
finite temperatures, the critical behaviour is essentially unchanged with
respect to the classical spherical model. At zero temperature, depending on
the parameter $g$, there is a quantum phase transition characterized by new
(quantum) critical exponents. Also, the introduction of quantum fluctuations
leads to a correction of the unphysical behaviour of the entropy at low temperatures.

We report an analysis of this version of the quantum mean spherical model in
the presence of competing interactions. We consider ferromagnetic
interactions, $J_{1}>0$, between pairs of first-neighbour sites along the $d$
directions of a hypercubic lattice, and antiferromagnetic interactions,
$J_{2}<0$, between second-neighbour sites along $m\leq d$ directions.
Classical versions of this model \cite{Kalok1972}\cite{Hornreich1975}%
\cite{Selke1977}\cite{Pisani1986}, as well as more elaborate mean spherical
models with competing interactions \cite{Richert1998}, have been studied by
several authors. For $m=1$, we regain a spherical analogue of the
Axial-Next-Nearest-neighbour Ising, or ANNNI, model \cite{Yokoi1981}%
\cite{Selke1988}, which is known to display a rich phase diagram, including a
Lifshitz point, in terms of the temperature $T$ and a parameter $p=-J_{2}%
/J_{1}$ that gauges the strength of the competing interactions. We then
analyse the $T-p-g$ phase diagram, for different values of $m$, in particular
the $T=0$ behaviour, and establish the critical dimensions and critical
exponents associated with this quantum model system. In the classical case,
$g=0$, we confirm a number of results, including a singularity of the critical
border at the Lifshitz point for $2<d<\left(  m+6\right)  /2$. In one
dimension, we derive analytic expressions for the decay of pair correlations,
and determine the region of modulated behaviour in the $T-p$ phase diagram.

\section{The quantum mean spherical model with competing interactions}

This problem can be treated either by a conventional reduction to a system of
coupled harmonic oscillators or by a judicious application of the method of
path integrals \cite{Nieuwenhuizen1995}\cite{Orland1988}\cite{AshokDas2006}.
Let us first use the representation in terms of harmonic oscillators. We then
introduce bosonic operators $a_{l}^{\dag}$ and $a_{l}$ to write%
\begin{equation}
S_{l}\equiv\frac{1}{\sqrt{2}}\left(  \frac{g}{2\mu}\right)  ^{1/4}\left(
a_{l}+a_{l}^{\dag}\right) \label{meq5}%
\end{equation}
and
\begin{equation}
P_{l}\equiv-\frac{i}{\sqrt{2}}\left(  \frac{2\mu}{g}\right)  ^{1/4}\left(
a_{l}-a_{l}^{\dag}\right)  ,\label{meq6}%
\end{equation}
where we have omitted the vector notation. We now assume periodic boundary
conditions, and change to a Fourier representation,%
\begin{equation}
a_{l}=\frac{1}{N^{d/2}}\sum_{q}a_{q}\exp\left(  iql\right)  ,
\end{equation}
where $a_{q}$ and $a_{q}^{\dag}$ are bosonic operators,%
\begin{equation}
\lbrack a_{q},a_{q^{\prime}}]=0,\quad\lbrack a_{q}^{\dag},a_{q^{\prime}}%
^{\dag}]=0,\quad\lbrack a_{q},a_{q^{\prime}}^{\dag}]=\delta_{q,q^{\prime}},
\end{equation}
and the sum is over the $d$-dimensional vectors $q$ belonging to the first
Brillouin zone. We then write the quantum quadratic form (\ref{hquantum}) in
the Fourier representation,
\begin{align}
\overline{\mathcal{H}} &  =(2g\mu)^{1/2}\sum_{q}\left[  1-\frac{\hat
{J}(q\mathbf{)}}{4\mu}\right]  a_{q}^{\dag}a_{q}-\frac{1}{2}(2g\mu)^{1/2}%
\sum_{q}\frac{\hat{J}(q)}{4\mu}\left(  a_{q}a_{-q}+a_{q}^{\dag}a_{-q}^{\dag
}\right) \nonumber\\
&  -H\left(  \frac{N^{d}}{2}\right)  ^{1/2}\left(  \frac{g}{2\mu}\right)
^{1/4}\left(  a_{0}+a_{0}^{\dag}\right)  +\frac{N^{d}}{2}\left(  2g\mu\right)
^{1/2},\label{hbarq}%
\end{align}
with%
\begin{equation}
\hat{J}(\overrightarrow{q}\mathbf{)=}%
{\displaystyle\sum\limits_{\overrightarrow{h}}}
J\left(  \left\vert \overrightarrow{h}\right\vert \right)  \exp\left(
i\overrightarrow{q}.\overrightarrow{h}\right)  ,\label{jq}%
\end{equation}
where the sum runs over all lattice vectors. The final diagonalization of this
quadratic form comes from the introduction of new bosonic operators, $c_{q}$
and $c_{q}^{\dag}$, according to a well-known Bogoliubov transformation. We
then have
\begin{equation}
\overline{\mathcal{H}}=\sum_{q}w\left(  q\right)  \left(  c_{q}^{\dag}%
c_{q}+\frac{1}{2}\right)  -\frac{N^{d}H^{2}}{4\left[  \mu-\frac{\hat{J}(0)}%
{2}\right]  },\label{diagonal}%
\end{equation}
where%
\begin{equation}
\left[  w\left(  q\right)  \right]  ^{2}\equiv(2g\mu)\left[  1-\frac{\hat
{J}(q)}{2\mu}\right]  ,\label{freq}%
\end{equation}
which requires that the chemical potential $\mu$ should be larger than a
certain limiting critical value $\mu_{c}$,%
\begin{equation}
\mu>\mu_{c}=\max\limits_{\overrightarrow{q}}\frac{1}{2}\hat{J}(\overrightarrow
{q})=\frac{1}{2}\hat{J}(\overrightarrow{q}_{c}).\label{mucrit}%
\end{equation}

In analogy with a system of harmonic oscillators, we then write the partition
function%
\begin{equation}
Z_{N}(\beta,H,\mu)=\exp\left[  \frac{\beta N^{d}H^{2}}{4\left(  \mu-\frac
{\hat{J}(0)}{2}\right)  }\right]  \prod_{q}\left[  2\sinh\frac{1}{2}\beta
w\left(  q\right)  \right]  ^{-1},\label{partfunc}%
\end{equation}
and the free energy per site,%
\begin{equation}
f(\beta,H,\mu)=-\frac{H^{2}}{4\left[  \mu-\frac{\hat{J}(0)}{2}\right]  }%
+\frac{1}{\beta}\lim_{N\rightarrow\infty}\frac{1}{N^{d}}\sum_{q}\ln\left[
2\sinh\frac{1}{2}\beta w\left(  q\right)  \right]  ,\label{energy}%
\end{equation}
with $\mu>\mu_{c}=\hat{J}(\overrightarrow{q}_{c})/2$. The spherical
constraint, given by Eq. (\ref{sphconstr}), from which we determine the
chemical potential $\mu$, is written as%
\begin{equation}
1=\frac{H^{2}}{4\left[  \mu-\frac{\hat{J}(0)}{2}\right]  ^{2}}+\lim
_{N\rightarrow\infty}\frac{1}{N^{d}}\sum_{q}\frac{g}{2w\left(  q\right)
}\coth\left[  \frac{1}{2}\beta w\left(  q\right)  \right]  .\label{constraint}%
\end{equation}
In the classical limit, $g\rightarrow0$, we regain most of the well-known
results for the mean spherical model. The classical limit of this quantum free
energy, however, includes an extra term of the form $\ln(\beta g)$, which
corrects the classical behaviour at low temperatures. The particular limit
$\hat{J}(q)\rightarrow0$ corresponds to free quantum (spherical) rotors, with
a finite energy gap, in contrast to the usual Heisenberg-Dirac spins.

These expressions also come from a straightforward application of the path
integral formalism, which has been widely used to treat quantum statistical
problems \cite{Nieuwenhuizen1995}\cite{Orland1988}. It is then interesting to
write the Lagrangian associated with this problem,%
\begin{equation}
\mathcal{L}=\frac{1}{2g}\sum_{l}\dot{S}_{l}^{2}+\frac{1}{2}\sum_{k,l}%
J_{k,l}S_{k}S_{l}+H\sum_{l}S_{l}-\mu\sum_{l}S_{l}^{2}.\label{lquantum}%
\end{equation}
In the imaginary time formalism, with $t\rightarrow-i\tau$, the partition
function is written as%
\begin{align}
Z &  =\int\left(  \prod_{l}\mathcal{D}S_{l}(\tau)\right)  \exp\left\{
\int_{0}^{\beta}d\tau\left[  -\frac{1}{2g}\sum_{l}\left(  \frac{\partial
S_{l}(\tau)}{\partial\tau}\right)  ^{2}\right.  \right. \nonumber\\
&  +\left.  \left.  \frac{1}{2}\sum_{l,l^{\prime}}J_{l,l^{\prime}}S_{l}%
(\tau)S_{l^{\prime}}(\tau)+H\sum_{l}S_{l}(\tau)-\mu\sum_{l}S_{l}^{2}%
(\tau)\right]  \right\}  ,
\end{align}
where the first integral includes periodic conditions, $S_{l}(0)=S_{l}(\beta
)$. We now introduce the Fourier transformation,
\begin{equation}
S_{l}(\tau)=\left(  \frac{\beta}{N}\right)  ^{\frac{1}{2}}\sum_{q,w}%
\exp\left[  i(ql+\tau w)\right]  S_{q}(w),
\end{equation}
where the vector $q$ belongs to the first Brillouin zone, and $w\equiv
w_{n}=(2n\pi)/\beta$, with integer $n$, is a Matsubara frequency. We then
write the partition function%
\begin{align}
Z &  =\exp\left(  \beta N\mu\right)  \int\left(  \prod_{n=-\infty}^{\infty
}\prod_{q}dS_{q}(w_{n})\right)  \exp\left\{  \left(  \beta N^{d}\right)
^{\frac{1}{2}}\beta S_{0}(0)H\right. \nonumber\\
&  +\left.  \sum_{q,n}\left[  -\frac{\beta^{2}w_{n}^{2}}{2g}+\frac{1}{2}%
\beta^{2}\hat{J}(q)-\beta^{2}\mu\right]  S_{q}(w_{n})S_{-q}(-w_{n})\right\}  ,
\end{align}
where $\hat{J}(q)$ is given by Eq. (\ref{jq}), and we have omitted the vector
notation. If we calculate the Gaussian integrals, and use the identity%
\begin{equation}
\prod_{n=1}^{\infty}\left[  1+\left(  \frac{w\beta}{n\pi}\right)  ^{2}\right]
=\frac{\sinh w\beta}{w\beta},
\end{equation}
it is straightforward to regain the partition function given by Eq.
(\ref{partfunc}).

This general solution works for all forms of distance-dependent interactions,
$J_{\overrightarrow{k},\overrightarrow{l}}=J\left(  \left\vert \overrightarrow
{k}-\overrightarrow{l}\right\vert \right)  $. We now consider ferromagnetic
interactions, $J_{1}>0$, between pairs of first-neighbour sites along the $d$
directions of a hypercubic lattice, and antiferromagnetic interactions,
$J_{2}<0$, between second-neighbour sites along $m\leq d$ directions. The
Fourier transform of the exchange interactions is given by
\begin{equation}
\hat{J}(\overrightarrow{q})=2J_{1}\sum_{j=1}^{d}\cos q_{j}+2J_{2}\sum
_{j=1}^{m}\cos2q_{j},\label{jfourier}%
\end{equation}
where $\overrightarrow{q}=(q_{1},q_{2},...,q_{d})$ is a wave vector in
suitable (dimensionless) units. The maximum of $\hat{J}(\overrightarrow{q})$
depends on the ratio $p=-J_{2}/J_{1}$. If $p\leq1/4$, the maximum is located
at the critical value $\overrightarrow{q}_{c}=0$, as in the simple
ferromagnetic case. If $p>1/4$, the maximum of $\hat{J}(\overrightarrow{q})$
is given by the critical vector%
\begin{equation}
\overrightarrow{q}_{c}=\left(  q_{c1},q_{c2},...,q_{cm},0,...0\right)  ,
\end{equation}
where%
\begin{equation}
q_{c1}=q_{c2}=...=q_{cm}=\cos^{-1}\frac{1}{4p}.
\end{equation}
The special case $p=1/4$ corresponds to a Lifshitz point of degree $m$.

\section{Phase diagrams and critical behaviour}

In zero field, $H=0$, the paramagnetic critical boundary in the $T-p-g$ space
comes from the spherical constraint, given by Eq. (\ref{constraint}),
supplemented by the critical limit of the chemical potential, $\mu=\mu_{c}$,
given by Eq. (\ref{mucrit}). We then write%
\begin{equation}
1=\lim_{N\rightarrow\infty}\frac{1}{N^{d}}\sum_{q}\frac{g}{2w_{c}\left(
q\right)  }\coth\left[  \frac{1}{2}\beta w_{c}\left(  q\right)  \right]
,\label{constraint2}%
\end{equation}
where%
\begin{equation}
w_{c}\left(  q\right)  =g^{1/2}\left[  \hat{J}(\overrightarrow{q}_{c})-\hat
{J}(\overrightarrow{q})\right]  ^{1/2}.\label{freq2}%
\end{equation}

In the classical limit, $g\rightarrow0$, we have%
\begin{equation}
1=\lim_{N\rightarrow\infty}\frac{1}{N^{d}}\sum_{q}\frac{1}{\beta\left[
\hat{J}(\overrightarrow{q}_{c})-\hat{J}(\overrightarrow{q})\right]  },
\end{equation}
from which we obtain the critical temperature as function of $p$, for all
values of $d$ and $m$,%
\begin{equation}
\frac{k_{B}T_{c}}{2J_{1}}=\frac{1}{I\left(  p,d,m\right)  },
\end{equation}
with
\begin{equation}
I\left(  p,d,m\right)  =\frac{1}{\left(  2\pi\right)  ^{d}}\int d^{d}q\frac
{1}{{\sum\limits_{j=1}^{d}}\left(  1-\cos q_{j}\right)  -p{\sum\limits_{j=1}%
^{m}}\left(  1-\cos2q_{j}\right)  },\label{i1}%
\end{equation}
for $p<1/4$, and
\begin{equation}
I\left(  p,d,m\right)  =\frac{1}{\left(  2\pi\right)  ^{d}}\int d^{d}q\frac
{1}{{\sum\limits_{j=1}^{d}}\left(  1-\cos q_{j}\right)  -p{\sum\limits_{j=1}%
^{m}}\left(  \frac{1}{p}-\frac{1}{8p^{2}}-1-\cos2q_{j}\right)  },\label{i2}%
\end{equation}
for $p>1/4$.

There is a long history associated with the calculations of similar lattice
Green functions \cite{Joyce1972}\cite{Zucker2011}. From the identity%
\begin{equation}
\frac{1}{a^{n}}=\frac{1}{\left(  n-1\right)  !}\int_{0}^{\infty}%
dx\,x^{n-1}\,\exp\left(  -ax\right)  ,\label{identity1}%
\end{equation}
where $n$ is an integer, and $a>0$, we write integral representations for
$I\left(  p,d,m\right)  $, which are convenient to carry out an asymptotic
analysis. For $d\leq2$, the divergence of these integrals indicate that
$T_{c}=0$ for all values of $m$ and $p\neq1/4$. Also, we have $T_{c}>0$ for
$d>2$, and for all values of $m$ and $p\neq1/4$. In particular, at the
Lifshitz point, $T_{c}>0$ for $d>(m+4)/2$. We now consider the graphs of
$T_{c}=T_{c}\left(  p\right)  $ versus the parameter $p$. It is easy to write
an expression for $dT_{c}/dp$, for $p<1/4$ and $p>1/4$, and to show that there
is common tangent at the Lifshitz point, $p=1/4$. For example, for $d=3 $ and
$m=1$, we have%
\[
\left.  \frac{\partial}{\partial p}\left(  \frac{k_{B}T_{c}}{2J_{1}}\right)
\right\vert _{p=1/4}=-\frac{1}{\left[  I\left(  \frac{1}{4},3,1\right)
\right]  ^{2}}\widetilde{I},
\]
with%
\begin{equation}
\widetilde{I}=\frac{1}{4\left(  2\pi\right)  ^{3}}\int d^{3}q\frac
{1-\cos2q_{1}}{\left(  \frac{11}{4}-\cos q_{1}-\cos q_{2}-\cos q_{3}%
+\cos2q_{1}\right)  ^{2}}.
\end{equation}
We then use the identity (\ref{identity1}), with $n=2$, to write an integral
representation from which it is easy to show that the common tangent at the
Lifshitz point, $p=1/4$, is infinite for (i) $d<2$, and (ii) $d>2$, with
$d<(m+6)/2$, which includes the analogue of the ANNNI model ($d=3$ and $m=1$).
In the numerically obtained graphs of figure 1, we sketch typical profiles of
the critical line $T_{c}=T_{c}\left(  p\right)  $ for dimensions $d=3$ and
$d=4$, and $m=1$. Note the smooth behaviour of this paramagnetic border for
$d=4$ (and $m=1$). The scale of this figure, however, is not enough to show
the sharp singularity at the Lifshitz point for $d=3$ (and $m=1$), as pointed
out in a sketch by Hornreich \cite{Hornreich1980}.%

\begin{figure}
[ptb]
\begin{center}
\includegraphics[
height=2.802in,
width=3.6313in
]%
{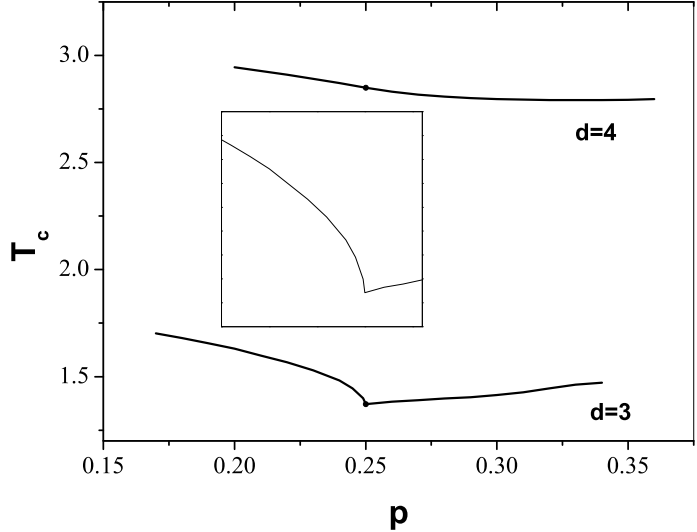}%
\caption{Classical $T-p$ phase diagrams, for dimensions $d=3$ and $d=4$, near
a Lifshitz point ($p=1/4$), with $m=1$. The critical line separates ordered
and disordered phases. Along the critical line, $\vec{q}_{c}=0$ for $p\leq
1/4$, and $\vec{q}_{c}\neq0$ \ for $p>1/4$. The inset shows a magnification in
order to emphasize the singular behavior of the paramagnetic border near the
Lifshitz point in three-dimensions.}%
\end{center}
\end{figure}

In the zero-temperature limit, $T\rightarrow0$, Eq. (\ref{constraint2}) can be
written as%
\[
1=\frac{g^{1/2}}{2}\lim_{N\rightarrow\infty}\frac{1}{N^{d}}\sum_{q}\frac
{1}{\left[  \hat{J}(\overrightarrow{q}_{c})-\hat{J}(\overrightarrow
{q})\right]  ^{1/2}},
\]
from which we obtain the critical quantum parameter, $g_{c}$, as a function of
$p$, for all values of $d$ and $m$,%
\begin{equation}
\frac{g_{c}}{2J_{1}}=\left[  \frac{2}{I_{Q}\left(  p,d,m\right)  }\right]
^{2},
\end{equation}
where%
\begin{equation}
I_{Q}\left(  p,d,m\right)  =\frac{1}{\left(  2\pi\right)  ^{d}}\int
d^{d}q\frac{1}{\left[  {\sum\limits_{j=1}^{d}}\left(  1-\cos q_{j}\right)
-p{\sum\limits_{j=1}^{m}}\left(  1-\cos2q_{j}\right)  \right]  ^{1/2}},
\end{equation}
for $p<1/4$, and%
\begin{equation}
I_{Q}\left(  p,d,m\right)  =\frac{1}{\left(  2\pi\right)  ^{d}}\int
d^{d}q\frac{1}{\left[  {\sum\limits_{j=1}^{d}}\left(  1-\cos q_{j}\right)
-p{\sum\limits_{j=1}^{m}}\left(  \frac{1}{p}-\frac{1}{8p^{2}}-1-\cos
2q_{j}\right)  \right]  ^{1/2}},
\end{equation}
for $p>1/4$. We now use an analytic continuation of the identity
(\ref{identity1}), for non integer values of $n$. From a similar analysis of
convergence of these expressions, it is easy to show that there is a quantum
phase transition ($g_{c}\neq0$) for $d\geq2$, independent of the value of $m$.
In particular, there is a common derivative $\partial g_{c}/\partial p$ at the
Lifshitz point, $p=1/4$, with a singularity for $2\leq d<(m+4)/2$ (and a
smooth behaviour for $d=3$ and $m=1$). In figure 2, we sketch typical profiles
of this critical line in the $g-p$ plane for dimensions $d=2$ and $d=3$, and
$m=1$.%

\begin{figure}
[ptb]
\begin{center}
\includegraphics[
height=2.7899in,
width=3.3797in
]%
{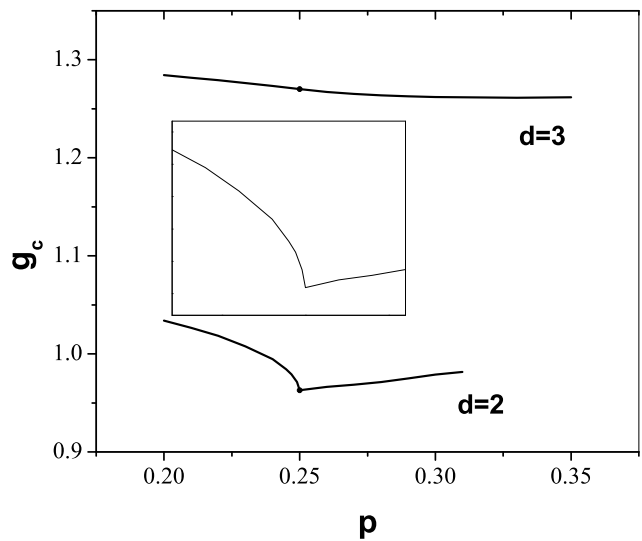}%
\caption{Quantum phase diagram ($T=0$) near a Lifshitz point with dimensions
$d=2$ and $d=3$, and $m=1$. Along the critical line, $g_{c}=g_{c}\left(
p\right)  $, we have $\vec{q}_{c}=0$ for $p\leq1/4$, and $\vec{q}_{c}\neq0$
\ for $p>1/4$. The inset shows a magnification of the border near $p=1/4$ in
two dimensions.}%
\end{center}
\end{figure}

The critical behaviour in zero field, $H=0$, comes from an asymptotic analysis
of the spherical constraint in the neighbourhood of the transition. At finite
temperatures, $T\neq0$, in the limit $\mu\rightarrow\hat{J}(\mathbf{q})/2$, we
have%
\begin{equation}
\frac{1}{N^{d}}\sum_{q}\frac{g}{2w\left(  q\right)  }\coth\frac{1}{2}\beta
w\left(  q\right)  =\frac{1}{12}g\beta+\frac{1}{N^{d}}\sum_{q}\frac{g}{\beta
w^{2}\left(  q\right)  }+\mathcal{O}[w^{2}\left(  q\right)  ].\label{sumfin}%
\end{equation}
At zero temperature, $T=0$, we have%
\begin{equation}
\frac{1}{N^{d}}\sum_{q}\frac{g}{2w\left(  q\right)  }\coth\frac{1}{2}\beta
w\left(  q\right)  =\frac{1}{N^{d}}\sum_{q}\frac{g}{2w\left(  q\right)
}.\label{sumzero}%
\end{equation}
We now expand $\hat{J}(q)$ as a Taylor series about $\overrightarrow
{q}=\overrightarrow{q}_{c}$, in the classical and quantum cases. Although
$\overrightarrow{q}_{c}$ depends on the parameter $p$, the convergence of the
sums in the right-hand side of Eqs. (\ref{sumfin}) and (\ref{sumzero}) does
not depend on $p$, for $p\neq1/4$. For finite temperatures, $T\neq0$, the sum
converges if $d>2$ (which determines the lower critical dimension of the
classical case), regardless of the value of $m$. At $T=0$, the sum converges
for $d>1$, which leads to the lower critical dimension of the quantum case.

Let us consider some special situations.

\subsection{Critical behaviour at finite temperatures and $p\neq1/4$}

In zero field, $H=0$, at finite temperatures, $T\neq0$, and for $p\neq1/4$, we
use Eq. (\ref{sumfin}) and perform a Taylor expansion about $T=T_{c}$ and
$\mu=\mu_{c}$. We then have the asymptotic expression%
\begin{equation}
0\approx\frac{(T-T_{c})}{2}\int d^{d}q~\frac{1}{\mu_{c}-\frac{\hat{J}(q)}{2}%
}-(\mu-\mu_{c})~T_{c}\int d^{d}q~\frac{1}{\left(  \mu_{c}-\frac{\hat{J}(q)}%
{2}\right)  ^{2}}.\label{Tfinite}%
\end{equation}
We now expand the integrands about $\overrightarrow{q}=\overrightarrow{q}_{c}
$, and obtain{%
\begin{equation}
(\mu-\mu_{c})\sim\left\{
\begin{array}
[c]{cc}%
\tau, & \text{for}~d>4\\
\frac{\tau}{\ln\tau}, & \text{for}~d=4\\
\tau^{2}, & \text{for}~d=3
\end{array}
\right.  ,\label{expclassical}%
\end{equation}
where $\tau=(T-T_{c})/T_{c}$ gives the distance from the classical critical
point. From these equations, supplemented by standard scaling considerations,
it is possible to calculate the usual critical exponents associated with the
classical spherical model. Although }$\overrightarrow{q}_{c}${\ assumes
different values for $p<1/4$ and $p>1/4$, the asymptotic behaviour is the
same, regardless of the value of }${m}${.}

\subsection{Critical behaviour at $T=0$ and $p\neq1/4$}

In analogy with the calculations for finite temperatures, we write{%
\begin{equation}
0\approx(g-g_{c})\int d^{d}q\frac{1}{\left(  \mu_{c}-\frac{\hat{J}%
(\mathbf{q})}{2}\right)  ^{1/2}}-(\mu-\mu_{c})g_{c}\int d^{d}q\frac{1}{\left(
\mu_{c}-\frac{\hat{J}(\mathbf{q})}{2}\right)  ^{3/2}}.\label{Tzero}%
\end{equation}
From an expansion about }$\overrightarrow{q}=\overrightarrow{q}_{c}$, we
obtain the asymptotic behaviour{\
\begin{equation}
(\mu-\mu_{c})\sim\left\{
\begin{array}
[c]{cc}%
\delta, & \text{for}~d>3\\
\frac{\delta}{\ln\delta}, & \text{for}~d=3\\
\delta^{2}, & \text{for}~d=2
\end{array}
\right.  ,\label{quantumexp}%
\end{equation}
where $\delta=(g-g_{c})/g_{c}$ gives the distance from the quantum critical
point. As in the case of finite temperatures, Eq. (\ref{quantumexp}) holds for
$p\neq1/4$ and any value of $m\leq d$. It is easy to use scaling arguments in
order to obtain the (quantum) critical exponents for $p\neq1/4$. With the
necessary reinterpretations, and although critical dimensions are different,
these values are in agreement with results of Vojta for the quantum
ferromagnetic case \cite{Vojta1996}.}

\subsection{Critical behaviour for $p=1/4$}

At the Lifshitz point, $p=1/4$, the maximum of $\widehat{J}\left(
\overrightarrow{q}\right)  $ is still given by $\overrightarrow{q}%
_{c}=(0,0,0,\cdots,0)$, but the second derivative vanishes along the direction
of competition. We then have to consider the quartic term in the expansion of
$\widehat{J}\left(  \overrightarrow{q}\right)  $ about $\overrightarrow
{q}=\overrightarrow{q}_{c}$.

For $T\neq0$, the first and second integrals of Eq. (\ref{Tfinite}) exist for
$d>(m+4)/2$ and for $d>(m+8)/2$, respectively. We then have%
\begin{equation}
(\mu-\mu_{c})\sim\left\{
\begin{array}
[c]{cc}%
\tau, & d>(m+8)/2,\\
{\tau}/{\ln\tau}, & d=(m+8)/2,\\
{\tau}^{3/2}, & d=(m+7)/2,\\
\tau^{2}, & d=(m+6)/2,\\
\tau^{5/2}, & d=(m+5)/2,
\end{array}
\right.  ,\label{classicalexp1}%
\end{equation}
where $\tau=(T-T_{c})/T_{c}$. From these asymptotic results, it is possible to
obtain all the classical critical exponents.

At $T=0$, the first integral in Eq. (\ref{Tzero}) exists for $d>(m+2)/2$, and
the second integral for $d>(m+6)/2$. We then have%
\begin{equation}
(\mu-\mu_{c})\sim\left\{
\begin{array}
[c]{cc}%
\delta, & d>(m+6)/2,\\
{\delta}/{\ln\delta}, & d=(m+6)/2,\\
{\delta}^{3/2}, & d=(m+5)/2,\\
\delta^{2}, & d=(m+4)/2,\\
\delta^{5/2}, & d=(m+3)/2,
\end{array}
\right.  ,\label{quantumexp1}%
\end{equation}
where $\delta=(g-g_{c})/g_{c}$. In conclusion, we have the same values for
either classical or quantum exponents. As in the case of $p\neq1/4$, the only
difference is the critical dimension. According to an old conjecture about
quantum critical behaviour, the quantum values of the critical exponents are
given by the corresponding values of the classical version of the system in
$d+z$ dimensions, where $z$ is a dynamical critical exponent. For $m<d$ the
dynamical critical exponent is $z=1$ (anisotropic case); for $m=d$, it assumes
the value $z=2$ (isotropic case). It should be mentioned that Hamiltonian
formulations of the mean spherical model have been considered by some authors
\cite{Srednicki1979}.

\subsection{Decay of pair correlations}

Let us consider the system in a site-dependent field,%
\begin{equation}
\mathcal{H}=-\sum_{\left(  \overrightarrow{k},\overrightarrow{l}\right)
}J_{\overrightarrow{k},\overrightarrow{l}}~S_{\overrightarrow{k}%
}S_{\overrightarrow{l}}-\sum_{\overrightarrow{l}}H_{\overrightarrow{l}%
}\,S_{\overrightarrow{l}},
\end{equation}
and write the partition function%
\begin{equation}
Z_{N}(\beta,H,\mu)=\prod_{q}\exp\left[  \frac{\beta N^{d}\widehat{H}\left(
q\right)  \widehat{H}\left(  -q\right)  }{4\left(  \mu-\frac{\hat{J}(q)}%
{2}\right)  }\right]  \left[  2\sinh\frac{1}{2}\beta w\left(  q\right)
\right]  ^{-1},\label{partfunc2}%
\end{equation}
where $\widehat{H}\left(  q\right)  $ is the Fourier transform of
$H_{\overrightarrow{l}}$, and we are omitting the vector notation. We then
have%
\begin{equation}
\left\langle S_{q}S_{-q}\right\rangle _{N}=\frac{4}{\left(  \beta
N^{d}\right)  ^{2}}\left.  \frac{\delta^{2}\ln Z}{\delta\widehat{H}\left(
q\right)  \delta\widehat{H}\left(  -q\right)  }\right\vert _{\widehat
{H}\left(  q\right)  =\widehat{H}\left(  -q\right)  =0}=\frac{2g}{\beta N^{d}%
}\frac{1}{\left[  w\left(  q\right)  \right]  ^{2}},
\end{equation}
from which we obtain the pair correlations in real space,%
\begin{equation}
\left\langle S_{r}S_{r+h}\right\rangle _{N}=\frac{1}{\beta N^{d}}\sum_{q}%
\frac{\exp\left(  iqh\right)  }{2\mu-\hat{J}(q)}.\label{corr}%
\end{equation}
As discussed by Pisani and collaborators \cite{Pisani1986}, the analysis of
$\left\langle S_{r}S_{r+h}\right\rangle $, for $d\geq3$, below the critical
temperature, leads to the introduction of a modulated order parameter, with
characteristic oscillations for $p>1/4$. A detailed analysis of the long-range
correlations at Lifshitz point has been published by Frachebourg and Henkel
\cite{Frachebourg1993}. There are also some investigations of the interplay
between competing interactions and the decay of correlations
\cite{Chakrabarty2011}.

We now show that these oscillations of the pair correlations in terms of
distance are already present in the much simpler one-dimensional case. In the
thermodynamic limit, Eq. (\ref{corr}) can be written as%
\begin{equation}
\left\langle S_{r}S_{r+h}\right\rangle =\frac{1}{2\pi\beta}\int_{-\pi}^{\pi
}dq~\frac{\exp\left(  iqh\right)  }{\left[  \mu-\cos q+p\cos2q\right]
},\label{c1}%
\end{equation}
with the spherical condition
\begin{equation}
1=\frac{1}{2\pi\beta}\int_{-\pi}^{\pi}dq~\frac{1}{\left[  \mu-\cos
q+p\cos2q\right]  },\label{esf1}%
\end{equation}
where $p>0$, $\mu$ and $\beta$ are written in units of $J_{1}>0$, with the
requirement that $\mu>\mu_{c}$. Therefore, $\mu>(1-p)$ for $p<1/4$, and
$\mu>[1/(8p)+p]$ for $p>1/4$. Eq. (\ref{c1}) can be rewritten as%
\begin{equation}
\left\langle S_{r},S_{r+h}\right\rangle =\frac{1}{\pi{\beta}ip}\oint
_{C}dw\frac{w^{h+1}}{w^{4}-\frac{1}{p}w^{3}+\frac{2{\mu}}{p}w^{2}-\frac{1}%
{p}w+1},
\end{equation}
where the contour $C$ is the unit circle and we assuming that $h>0$. The
fourth-order polynomial in the denominator is easily factorized, with two
roots, $w_{1}$ and $w_{2}$, inside the unit circle. After some straightforward
algebra, we have%
\begin{equation}
\left\langle S_{r}S_{r+h}\right\rangle =\frac{\cos\left[  h\theta_{1}+\left(
\theta_{1}-\alpha\right)  \right]  }{\cos\left(  \theta_{1}-\alpha\right)
}\exp\left[  h\ln|w_{1}|\right]  ,\label{c11}%
\end{equation}
where $\alpha$ is a real phase depending on $p$ and $\mu$, and%
\begin{equation}
w_{1}=\left\vert w_{1}\right\vert \exp\left(  i\theta_{1}\right)  =\frac
{1}{4p}(1-A-2B),
\end{equation}
with%
\begin{equation}
A=\sqrt{8p^{2}-8p\mu+1};\qquad B=\sqrt{\frac{1}{2}-2p\left(  p-\mu\right)
-\frac{1}{2}A},
\end{equation}
so that $\left\vert w_{1}\right\vert <1$, which leads to the well-known
exponential decay. The spherical condition (\ref{esf1}) can be used to
parametrically eliminate the chemical potential $\mu$, and write the
correlations in terms of $T$ and $p$. In particular, oscillations are
suppressed by $\theta_{1}=0$, which is equivalent to%
\begin{equation}
\frac{k_{B}T}{2J_{1}}=\frac{1}{8p}-2p,
\end{equation}
with the asymptotic value $T=0$ for $p=1/4$. In Fig. 3, we draw this border,
and indicate the region with oscillating correlations ($\theta_{1}\neq0$) in
the $T-p$ plane. From eq. (\ref{c11}), we can write expressions for a
modulation length, $L_{D}=2\pi/\theta_{1}$, and the correlation length,
$\xi=-1/\ln\left\vert w_{1}\right\vert $, in terms of temperature and the
competition parameter $p$, which is a useful information to investigate the
growth of modulated domains \cite{Chakrabarty2011}. A similar behaviour has
been found by Stephenson \cite{Stephenson1970} in a calculation for the ANNNI chain.%

\begin{figure}
[ptb]
\begin{center}
\includegraphics[
height=2.802in,
width=3.3512in
]%
{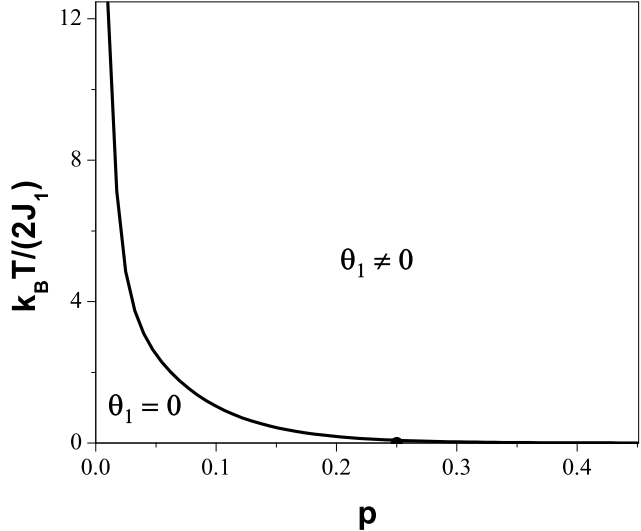}%
\caption{Regions of the $T-p$ plane of the one-dimensional model with simple
exponential decay ($\theta_{1}=0$) and oscillatory exponential decay
($\theta_{1}\neq0$) of the pair correlations.}%
\end{center}
\end{figure}

\section{Conclusions}

We report an analysis of the phase diagram of a quantum mean spherical model
in terms of temperature $T$, a quantum parameter $g$, and the ratio
$p=-J_{2}/J_{1}$, where $J_{1}>0$ is a ferromagnetic interaction between
first-neighbour sites along the $d$ directions of a hypercubic lattice, and
$J_{2}<0$ is associated with competing antiferromagnetic interactions between
second neighbours along $m\leq d$ directions. We regain a number of results
for the classical version of this model, including the topology of the
critical line in the $g=0$ space, with a singular behaviour at the Lifshitz
point, $p=1/4$, for $2<d<(m+6)/2$, which includes the case of the usual
analogue of the Axial-Next-Nearest-neighbour Ising, or ANNNI, model. We
consider in particular the $T=0$ phase diagram, which displays a quantum
Lifshitz point, at $p=1/4$. In the $g-p$ phase diagram, there is a critical
border, $g_{c}=g_{c}\left(  p\right)  $ for $d\geq2$, with a singularity at
the Lifshitz point if $d<(m+4)/2$. We establish upper and lower critical
dimensions and analyse the critical behaviour in the neighbourhood of the
Lifshitz point. In one dimension, we derive analytic expressions for the decay
of pair correlations, and determine the region of modulated behaviour in the
$T-p$ phase diagrams.\bigskip

\textbf{Acknowledgements}

We acknowledge the financial support of Brazilian agencies CAPES and CNPq.

\end{document}